\documentclass[aps,prd,twocolumn,superscriptaddress,bibnotes,longbibliography,preprintnumbers,floatfix,10pt]{revtex4-2}

\usepackage{graphicx}
\usepackage{xcolor}
\usepackage{amsmath,amsfonts,amsthm,amssymb}
\usepackage[colorlinks]{hyperref}
\usepackage{mathtools}
\usepackage{bm}
\usepackage{multirow}
\graphicspath{{./figures/}}

\newcommand{\orcid}[1]{\href{https://orcid.org/#1}{\includegraphics[width=10pt]{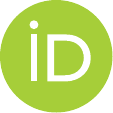}}}

\newcommand{\ror}[1]{\href{https://ror.org/#1}{\includegraphics[width=10pt]{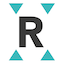}}}

\hypersetup{
    pdfnewwindow=true,      % links in new window
    colorlinks=true,       % false: boxed links; true: colored links
    linkcolor=violet,          % color of internal links
    citecolor=violet,        % color of links to bibliography
    filecolor=violet,      % color of file links
    urlcolor=violet        % color of external links
}

\begin{document}

\title{Measuring the Cosmic Ray Spectrum with Next Generation Neutrino Detectors}

\author{Stephan A. Meighen-Berger \orcid{0000-0001-6579-2000}\,}
\email{stephan.meighenberger@unimelb.edu.au}
\affiliation{School of Physics, The University of Melbourne, Victoria 3010, Australia \ror{01ej9dk98}}
\affiliation{Center for Cosmology and AstroParticle Physics (CCAPP), Ohio State University, 
Columbus, OH 43210, USA \ror{00rs6vg23}}

\author{Jayden L. Newstead \orcid{0000-0002-8704-3550}\,}
\email{jnewstead@unimelb.edu.au}
\affiliation{School of Physics, The University of Melbourne, Victoria 3010, Australia \ror{01ej9dk98}}
\affiliation{ARC Centre of Excellence for Dark Matter Particle Physics, School of Physics, The University of Melbourne, Victoria 3010, Australia \ror{01ej9dk98}}

\author{Louis E. Strigari \orcid{0000-0001-5672-6079}\,}
\email{strigari@tamu.edu}
\affiliation{Department of Physics and Astronomy, Texas A\&M University, College Station, TX 77843, USA \ror{01f5ytq51}}
\affiliation{George P. and Cynthia Woods Mitchell Institute for Fundamental Physics and Astronomy, Texas A\&M University, College Station, TX 7783, USA \ror{01f5ytq51}}

%\author{Yi Zhuang \orcid{0000-0002-7713-8724}\,}
%\email{yiz5@tamu.edu}
%\affiliation{George P. and Cynthia Woods Mitchell Institute for Fundamental Physics and Astronomy, Texas A\&M University, College Station, TX 7783, USA \ror{01f5ytq51}}

\date{\today}

\begin{abstract}
We investigate the capabilities of upcoming kiloton-scale neutrino detectors, such as Hyper-Kamiokande, in determining the primary cosmic ray spectrum. These detectors provide full-sky coverage and long-term monitoring, unlike traditional satellite and balloon experiments that measure cosmic ray flux at specific altitudes and locations. By analyzing the atmospheric neutrino flux generated by cosmic ray interactions, we demonstrate that future detectors can differentiate between various cosmic ray models with high statistical significance, even when accounting for uncertainties in neutrino cross sections and hadronic interactions. We introduce a technique for reconstructing the primary cosmic ray spectrum using neutrino measurements, which reduces the flux uncertainty from approximately 20\% to about 7\%. We then show that Hyper-K has the potential to increase sensitivity to neutrino oscillation parameters, such as $\sin^2\theta_{23}$, by a factor of 2. Our results highlight the complementary role of neutrino detectors in cosmic ray physics and their critical importance for precision measurements in particle astrophysics.
\end{abstract}

\maketitle

%%%%%%%%%%%%%%%%%%%%%%%%%%%%%%%%%%%%%%%%%%%%%%%%%%%%%%%%%%%%%%%%%%%%%%%%%%%%%%%%%%%%%%%%%%%
%%%%%%%%%%%%%%%%%%%%%%%%%%%%%%%%%%%%%%%%%%%%%%%%%%%%%%%%%%%%%%%%%%%%%%%%%%%%%%%%%%%%%%%%%%%

\section{Introduction}\label{sec:intro}

\begin{figure}[t]
\centering
\includegraphics[width=0.9\columnwidth]{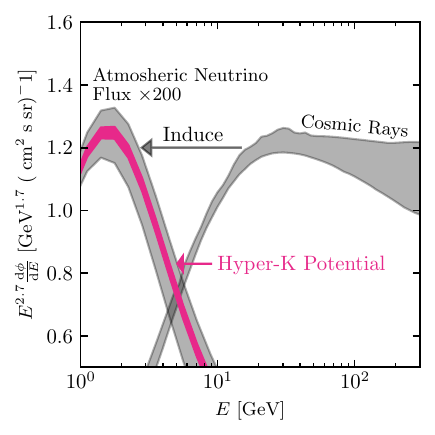}
\vspace{-1\baselineskip}% Reduce image height
\caption{A sketch of where differences in the primary flux will appear in neutrino data. Shown (in black) are the neutrino fluxes produced when injecting PAMELA, BESS, AMS, or HKKM CR proton. The pink area denotes Hyper-K's predicted statistical precision when measuring the neutrino flux.}
\label{fig:sketch}
\end{figure}

Next-generation large-scale neutrino detectors such as JUNO~\cite{JUNO:2015zny}, Hyper-K~\cite{Hyper-Kamiokande:2018ofw}, and DUNE~\cite{DUNE:2020txw} are poised to significantly advance our understanding of neutrino sources in the $\sim$MeV energy range, including the Sun, the atmosphere, diffuse supernovae, and potentially individual supernova events. Similarly, upcoming large-volume dark matter detectors will achieve sensitivity to low-energy neutrinos through independent detection channels, enabling complementary studies of these sources~\cite{Aalbers:2022dzr}.

Among these, atmospheric neutrinos~\cite{NUSEX:1988jvu, Becker-Szendy:1992ory, Becker-Szendy:1992ory, Soudan2:2003qqa, Gonzalez-Garcia:2006xta, Super-Kamiokande:2010orq, IceCube:2010whx, IceCube:2012jwm} constitute a critical signal. A precise understanding of the low-energy atmospheric neutrino flux ($E_\nu< 100$ MeV) is crucial for multiple physics goals: probing neutrino oscillations ~\cite{Super-Kamiokande:1998kpq, KM3Net:2016zxf, Super-Kamiokande:2017edb, IceCube:2019dqi, Esteban:2020cvm, Meighen-Berger:2023xpr}, mass ordering~\cite{IceCube-Gen2:2019fet, KM3NeT:2021ozk}, flavor ratios~\cite{Kamiokande:1994sgx, Super-Kamiokande:1998wen}, and it acts as a background in the search for the diffuse supernova neutrino background (DSNB)~\cite{Horiuchi:2008jz, Beacom:2010kk, KamLAND:2011bnd, JUNO:2015zny, Super-Kamiokande:2021jaq}, proton decay~\cite{KamLAND:2008dgz, JUNO:2015zny, Super-Kamiokande:2016exg, Hyper-Kamiokande:2018ofw, DUNE:2020ypp} indirect~\cite{Olivares-DelCampo:2017feq, Arguelles:2019ouk, Super-Kamiokande:2020sgt, Bell:2022ycf} and in direct detection searches for dark matter~\cite{Strigari:2009bq, Billard:2013qya, OHare:2021utq}.

Atmospheric neutrinos originate from cosmic ray (CR) interactions in the Earth's atmosphere, which produce showers that generate neutrinos as a secondary component. A variety of satellite-, balloon-, and ground-based experiments have directly measured the primary CR proton and helium spectra, including AMS~\cite{AMS:2015tnn}, CREAM~\cite{Yoon:2011aa}, PAMELA~\cite{PAMELA:2011mvy}, BESS~\cite{Abe:2015mga}, helium: AMS~\cite{AMS:2015azc}, ATIC-2~\cite{Panov:2009iih}, BESS~\cite{Abe:2015mga}, CREAM~\cite{Yoon:2011aa}, PAMELA~\cite{PAMELA:2011mvy}, among others. These measurements are generally well-described by phenomenological and empirical cosmic ray source models, such as HillasGaisser~\cite{Gaisser:2011klf}, Gaisser-Stanev-Tilav~\cite{Gaisser:2013bla}, HKKM~\cite{Honda:2016qyr, honda2011improvement}. Atmospheric shower simulations, such as HKKM~\cite{Honda:2016qyr, honda2011improvement}, CORSIKA~\cite{Heck:1998vt}, and MCEq~\cite{Fedynitch:2015zma}, are then used to predict the atmospheric neutrino spectrum, for a given primary CR spectrum as input. These spectra are a critical input to \textit{any} neutrino analysis. While~\cite{Evans:2016obt} previously studied how one can combine different CR data sets to reduce the uncertainty on the neutrino flux, here we reverse the logic. 

% This was based on the GSHL model from ~\cite{Gaisser:2002jj}

With the increasing precision and scale of neutrino detectors, it is timely to investigate how measurements of atmospheric neutrinos can inform phenomenological models of the primary CR spectrum. At the low energies relevant for atmospheric neutrino production, two principal effects on the CR spectrum must be considered: solar modulation~\cite{2013SSRv..176..299M}, which alters the spectrum as particles traverse the heliosphere, and geomagnetic deflection, which imposes a rigidity cut-off that varies with location and direction at the Earth's surface.

Concerning the atmospheric neutrino flux, experiments such as Super-Kamiokande have provided high-statistics measurements~\cite{Super-Kamiokande:2023ahc}, constraining neutrino oscillation parameters and even hinting at time variability correlated with cosmic ray modulation. Recent studies have explored the impact of detector location~\cite{Zhuang:2021rsg} and solar cycle variations~\cite{Kelly:2023ugn} on the low-energy atmospheric neutrino flux, highlighting the sensitivity of future detectors to these effects, particularly at higher latitudes.

In this work we evaluate the potential for future experiments to precisely measure the atmospheric neutrino flux, and show how this flux may be used to measure the CR spectrum. Figure~\ref{fig:sketch} sketches out our general approach. We use experimental CR flux measurements (right grey shaded band) to predict the resulting neutrino flux (left grey shaded band) at Hyper-K. We then show that Hyper-K's statistical precision can differentiate between these injected fluxes (pink shaded band).

In Sections \ref{sec:cr} and \ref{sec:nu_shower} we define the primary CR and shower models, respectively, that we use in this analysis. In Section \ref{sec:potential}, using HyperK as an example, we show how atmospheric neutrinos measurements can directly inform models of primary CR spectra and CR shower physics. Then, in Section \ref{sec:oscillations}, we show how this improved knowledge of the models can be leveraged to improve atmospheric neutrino oscillation measurements. Lastly, in Section \ref{sec:conclusion} we close with an outlook and concluding remarks.

%%%%%%%%%%%%%%%%%%%%%%%%%%%%%%%%%%%%%%%%%%%%%%%%%%%%%%%%%%%%%%%%%%%%%%%%%%%%%%%%%%%%%%%%%%%
%%%%%%%%%%%%%%%%%%%%%%%%%%%%%%%%%%%%%%%%%%%%%%%%%%%%%%%%%%%%%%%%%%%%%%%%%%%%%%%%%%%%%%%%%%%

\section{Cosmic Rays}\label{sec:cr}
The atmospheric neutrino flux at energies $E_\nu\leq 100\;\mathrm{GeV}$ is primarily sourced by CRs with energies $E_\mathrm{CR}\in [1\;\mathrm{GeV}, 1\;\mathrm{TeV}]$. For simplicity, we assume these CR to be purely composed of protons. Although there is a subleading helium component ($\approx$10\% - 20\% as measured by AMS~\cite{AMS:2015azc}), we assume the neutrino spectrum produced from these showers is the same as that produced by protons. Some shower observables, such as the depth of shower maximum $X_\mathrm{max}$, the total muonic component of the shower $N_\mu$, and the characteristic width of the shower profile $L$, do differ between $p$ and $He$ initiated showers~\cite{Buitink:2021pkz, Flaggs:2023exc}, for simplicity we do not account for these differences as they do not affect our results. 

Here we define three primary CR models to illustrate the effects that they have on the final neutrino spectrum and the neutrino detectors' capabilities to distinguish between them. We will use the measurements by AMS~\cite{AMS:2015tnn} as a benchmark and compare the neutrino spectra when injecting PAMELA~\cite{PAMELA:2011mvy} data and the CR model used in HKKM~\cite{Honda:2016qyr, honda2011improvement}. For low energies ($E_\mathrm{CR} < 1$ TeV) HKKM uses a proton and helium spectra mainly based on AMS02~\cite{AMS:2015tnn}. Above 1 TeV the HKKM model is based on CREAM~\cite{Yoon:2011aa}, JACEE~\cite{Asakimori_1998}, and RUNJOB~\cite{RUNJOB:2000zce}.

%%%%%%%%%%%%%%%%%%%%%%%%%%%%%%%%%%%%%%%%%%%%%%%%%%%%%%%%%%%%%%%%%%%%%%%%%%%%%%%%%%%%%%%%%%%
%%%%%%%%%%%%%%%%%%%%%%%%%%%%%%%%%%%%%%%%%%%%%%%%%%%%%%%%%%%%%%%%%%%%%%%%%%%%%%%%%%%%%%%%%%%

\section{Neutrino Showers}\label{sec:nu_shower}

At the energies of interest, atmospheric neutrinos are primarily produced by sequential decays of pions and muons produced by CR interactions in the upper atmosphere~\cite{gaisser2016cosmic, Matthews:2005sd, Engel:2011zzb}. Here we ignore muons produced in electromagnetic showers due to their minimal contribution~\cite{PGEdwards_1985, Dembinski:2021szp, Cazon:2022msf}. We simulate the production of neutrinos in the atmosphere using the site-dependent solar-cycle averaged predictions of HKKM11~\cite{Honda:2016qyr}, which agree well with Super-K measurements~\cite{Super-Kamiokande:2015qek}. For energies below 0.15 GeV (where the HKKM11 predictions stop), we use predictions from~\cite{Zhuang:2021rsg}, which combines FLUKA~\cite{Battistoni:2005pd} and CORSIKA~\cite{Heck:1998vt} to make site-dependent low-energy flux predictions. 

To model the effects of different injected CR spectra and hadronic models, we use MCEq~\cite{Fedynitch:2015zma}, which is a cascade-equation~\cite{Lipari:1993hd, Bossard:2000jh, Bergmann:2006yz, gaisser2016cosmic} based simulation tool. We then use the resulting changes predicted by MCEq to rescale the HKKM11 and low-energy atmospheric fluxes. In MCEq we used the hadronic interaction models Sybill 2.3~\cite{Riehn:2017mfm, Riehn:2019jet}, EPOS-LHC~\cite{Pierog:2013ria}, QGSJET-II~\cite{Ostapchenko:2010vb}, and DPMJET-III~\cite{Roesler:2000he, Fedynitch:2015kcn} as well as the atmospheric models from CORSIKA (US Standard)~\cite{Heck:1998vt}, and NRLMSISE-00~\cite{2002JGRA..107.1468P} to assess the uncertainties from the underlying hadronic and atmospheric models. Note~\cite{Fedynitch:2022vty} has shown that the uncertainties from Hadronic interaction models can be pushed down to $\sim 5\%$ for energies $E_\nu \leq 10\;\mathrm{GeV}$. This uncertainty is primarily caused by uncertainties from $\pi$ and $K$ production in the atmosphere~\cite{Honda:2006qj}, while the kaon to pion ratio itself should be irrelevant~\cite{Gaisser:2011klf}.

%%%%%%%%%%%%%%%%%%%%%%%%%%%%%%%%%%%%%%%%%%%%%%%%%%%%%%%%%%%%%%%%%%%%%%%%%%%%%%%%%%%%%%%%%%%
%%%%%%%%%%%%%%%%%%%%%%%%%%%%%%%%%%%%%%%%%%%%%%%%%%%%%%%%%%%%%%%%%%%%%%%%%%%%%%%%%%%%%%%%%%%

\section{Measurement Potential}\label{sec:potential}

Hyper-K, the successor to Super-K~\cite{Super-Kamiokande:2002weg}, is a cylindrical water-based Cherenkov detector with a fiducial volume of 187 kton~\cite{Hyper-Kamiokande:2018ofw, Hyper-Kamiokande:2022smq}. With angular resolution $\sim 10^\circ$~\cite{Super-Kamiokande:2005mbp} for $p_{e,\mu}=10\;\mathrm{GeV}$, detection efficiency $> 80\%$~\cite{Super-Kamiokande:2015qek} for CC $\nu_\mu$ events, and excellent energy resolution $\leq 10\%$~\cite{Shiozawa:1999sd, Drakopoulou:2017apf, Super-Kamiokande:2019gzr}, this detector will have an order of magnitude more well-reconstructed data than previous generations of detectors for neutrino energies below 10 GeV.

Here we consider atmospheric neutrino CC scattering on oxygen for Hyper-K. To predict the number of atmospheric events in an energy bin $j$, we use
\begin{equation}
    \mathcal{N}^j = N_{t} \, \Delta t \int_{E_j}^{E_{j+1}} \mathrm{d}E_\nu \, 
    \frac{\mathrm{d}\Phi}{\mathrm{d}E_\nu}(E_\nu) \, \sigma(E_\nu)\epsilon (E_\nu),
\label{eq:yield}
\end{equation}
where $N_t$ is the number of target atoms (water), $\epsilon$ the detection efficiency (80\% for Hyper-K), and $\Delta t$ is the detector livetime.  For the exposure, we assume 10 years of livetime for Hyper-K. The neutrino-oxygen cross section, $\sigma$, is taken from {\tt GENIE 3.2.0} using tune G18\_10a\_02\_11b, which is based on a local Fermi-gas model and an empirical meson-exchange model~\cite{Andreopoulos:2009rq, Andreopoulos:2015wxa, GENIE:2021zuu}. Here, we use the neutrino-oxygen cross section as an approximation for the total neutrino-water cross section. For neutrino energies between 100 MeV and 10 GeV, we predict $\sim 62,000$ $\nu_e$ and $\sim 110,000$ $\nu_\mu$ events. agreeing well with the predictions of~\cite{Zhou:2023mou} when scaled to Super-K. \textit{This implies that Hyper-K's statistical uncertainty is below 1\%}. As a result, Hyper-K has the statistical power to distinguish between the different cosmic-ray models, which can differ by up to 20\%.

\begin{figure}[t]
\centering
\includegraphics[width=\columnwidth]{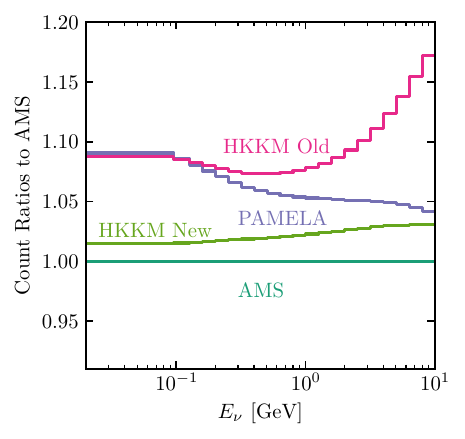}
\vspace{-2\baselineskip}% Reduce image height
\caption{The resulting neutrino counts compared to the AMS prediction when injecting PAMELA or HKKM data.}
\label{fig:fit_neutrino_ratio}
\end{figure}

Figure \ref{fig:fit_neutrino_ratio} shows the resulting neutrino count ratios (compared to an AMS injection) for the $\nu_e$ CC channel when injecting the different primary CR models and measurements. Note that the shape of the neutrino spectrum changes depending on energy for the different injections. \textit{This shows that a flat scaling cannot capture the energy-dependent uncertainty of the primary flux.} This will become particularly relevant once future experiments reach $\sim 5\%$ statistical precision.

To illustrate the effect of using different primary cosmic ray spectra we first perform a simple analysis which focuses on the resulting shape of the atmospheric neutrino spectrum. We define the test statistic, $q$, as a maximum likelihood ratio, where the Poissonian likelihood is maximized with respect to two nuisance parameters: the overall normalization (taken as a free parameter) and the CC cross section (taken to be Gaussian with 10\% uncertainty). We run the analysis for two distinct channels simultaneously: CC $\nu_e$ and CC $\nu_\mu$.

\begin{figure}[t]
\centering
\includegraphics[width=\columnwidth]{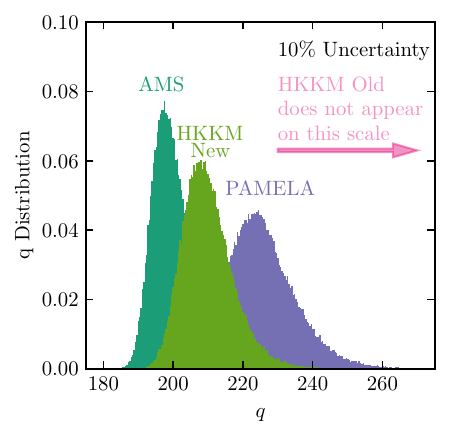}
\vspace{-2\baselineskip}% Reduce image height
\caption{Distribution of the test statistic for the primary CR models (holding the interaction model fixed) and accounting for a 10\% cross-section uncertainty in the CC oxygen channel.}
\label{fig:stat_q}
\end{figure}

Figure~\ref{fig:stat_q} shows the resulting distribution of the test statistic when injecting different primary cosmic ray fluxes. The differences arise for multiple reasons: The large difference between the old HKKM model and AMS data arises at high energies, where changes in the spectral shape of the cosmic ray flux were not accounted for. PAMELA measurements are consistently higher than AMS measurements and show the most significant differences at $\sim 700$ GeV. The new HKKM model interpolates between low-energy data (AMS) and higher-energy measurements (CREAM, JACEE, and RUNJOB). This causes the HKKM model to lie slightly above the AMS measurements starting at a few 100 GeV. Not shown are measurements by other experiments, such as older AMS measurements~\cite{ALPAT2002179}, ATIC-2~\cite{2009BRASP..73..564P}, and BESS-TeV~\cite{Haino:2004nq}. These differ more than PAMELA from AMS and would result in more significant changes in the test statistic.

From these distributions, we can extract the expected sensitivity of Hyper-K to the injected cosmic-ray fluxes using 10 years of data. For HKKM New, PAMELA, and HKKM Old, we predict sensitivities of $1.1\sigma$, $2.1\sigma$, and $\geq 5\sigma$, respectively, relative to the baseline fluxes using AMS data. By changing the energy range analyzed to $E_\nu\in [1\;\mathrm{GeV}, 10\;\mathrm{GeV}]$ we can slightly increase the sensitivity to the differences between AMS data and the model used in HKKM New to $1.4\sigma$. These results point to two facts: \textit{Hyper-K will be sensitive to the differences measured by cosmic-ray experiments} and \textit{Hyper-K will even be sensitive to the interpolation between low-energy and high-energy cosmic-ray data.}

Using the 10\% uncertainty on the neutrino-nucleus cross section assumed here, there is no sensitivity to the helium component of the cosmic-rays. Should the uncertainty be reduced to 3\%, Hyper-K would start to become sensitive to the measured helium flux. Note that this should be treated as a conservative statement, since we assume that helium and protons produce the same neutrino spectrum.

\section{Cosmic Ray Model}

Using the above results, we can then ask the question: \textit{Can Hyper-K use its data to improve the cosmic-ray model?}

To this end, we reverse the analysis, reconstructing the primary spectrum based on all-sky $\nu_e$ and $\nu_\mu$ measurements using the following procedure: First, we inject single proton lines at seven different energies and, following the same procedure as in the previous section, construct the resulting neutrino flux templates. We then use these seven neutrino spectra to fit the predicted Hyper-K all-sky measurement, binned as ten logarithmically spaced bins per decade in energy, using a $\chi^2$. In the fit, we include a systematic uncertainty on the neutrino-oxygen cross section of 10\%. Note that we are free to increase or decrease the number of injected lines (templates). However, increasing the template count increases the uncertainty for each proton line, since there are large degeneracies between the resulting neutrino spectra from each line. Additionally, decreasing the number of templates makes reproducing the neutrino spectrum's shape more challenging. 

\begin{figure}[t]
\centering
\includegraphics[width=\columnwidth]{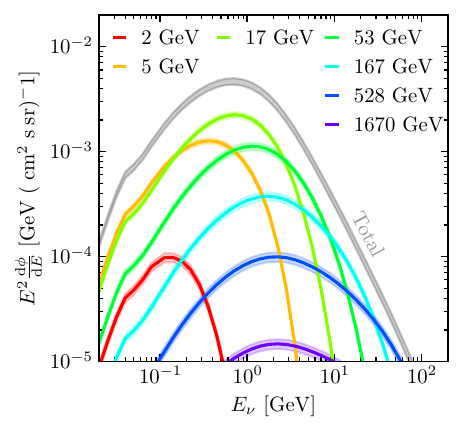}
\vspace{-2\baselineskip}% Reduce image height
\caption{Fitted neutrino spectrum, for discrete primary injections. Here, we fit 1900 kton-years worth of data. The shaded bands denote the uncertainty on each respective fit. Here we include a 10\% systematic uncertainty on the neutrino-oxygen cross section.}
\label{fig:fit_flux}
\end{figure}

Figure \ref{fig:fit_flux} shows the resulting fit. Each colored curve represents a neutrino flux template from a single injected CR proton line with a given energy. The resulting uncertainty on the reconstructed spectrum ranges from 5\% to 10\%, depending on the energy. \textit{The main benefit of this analysis is reducing the uncertainty on the neutrino fluxes from 15\% - 25\%~\cite{Super-Kamiokande:2017yvm} to 5\% - 10\%.} Interestingly, this is similar to the best-fit values of the uncertainty on the flux normalization Super-K finds: for $E_\nu < 1\;\mathrm{GeV}$: $14.3\%$ and for $E_\nu > 1\;\mathrm{GeV}$: $7.8\%$~\cite{Super-Kamiokande:2017yvm}.

\begin{figure}[t]
\centering
\includegraphics[width=\columnwidth]{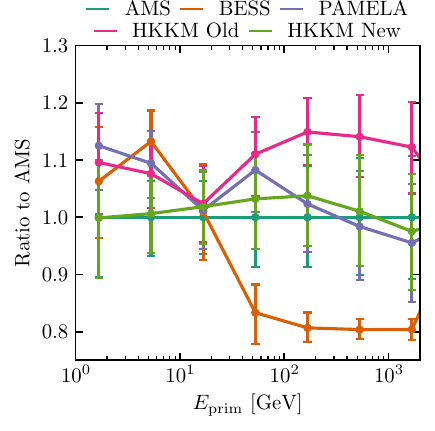}
\vspace{-2\baselineskip}% Reduce image height
\caption{The unfolded CR spectra based on predicted Hyper-K measurements for different injected fluxes.}
\label{fig:fit_CR}
\end{figure}

Figure \ref{fig:fit_CR} compares the fitted CR spectra for different injections to an AMS injection. Included are the resulting uncertainties on the fit of each CR energy bin. Of particular interest here is that the uncertainties and resulting fit values are \textit{energy dependent}. This is of particular interest to studies that probe specific neutrino energies, such as oscillation studies.

%%%%%%%%%%%%%%%%%%%%%%%%%%%%%%%%%%%%%%%%%%%%%%%%%%%%%%%%%%%%%%%%%%%%%%%%%%%%%%%%%%%%%%%%%%%
%%%%%%%%%%%%%%%%%%%%%%%%%%%%%%%%%%%%%%%%%%%%%%%%%%%%%%%%%%%%%%%%%%%%%%%%%%%%%%%%%%%%%%%%%%%
\section{Effects on oscillation studies}
\label{sec:oscillations} 

A reduction in the uncertainty of the cosmic ray spectrum will have important implications for measuring neutrino mixing parameters. Here we explore this and show the effects that reduced cosmic ray uncertainties would have on a simplified $\sin^2\theta_{23}$ study. This is meant to be considered a purely illustrative case, and may be extended with a more in-depth analysis. 

We first construct oscillograms using {\tt nuCraft}~\cite{Wallraff:2014qka}, and then fold them with the fitted neutrino spectra from the previous section. Using Equation~\ref{eq:yield}, we then predict the number of events per energy and angle bin, and compare the results for the different choice of neutrino parameters. Since we are comparing absolute counts in the analysis region between different $\sin^2\theta_{23}$ hypotheses, the result is susceptible to the uncertainty on the overall normalization of the neutrino flux caused by the cross-section and primary model.

\begin{figure}[t]
\centering
\includegraphics[width=\columnwidth]{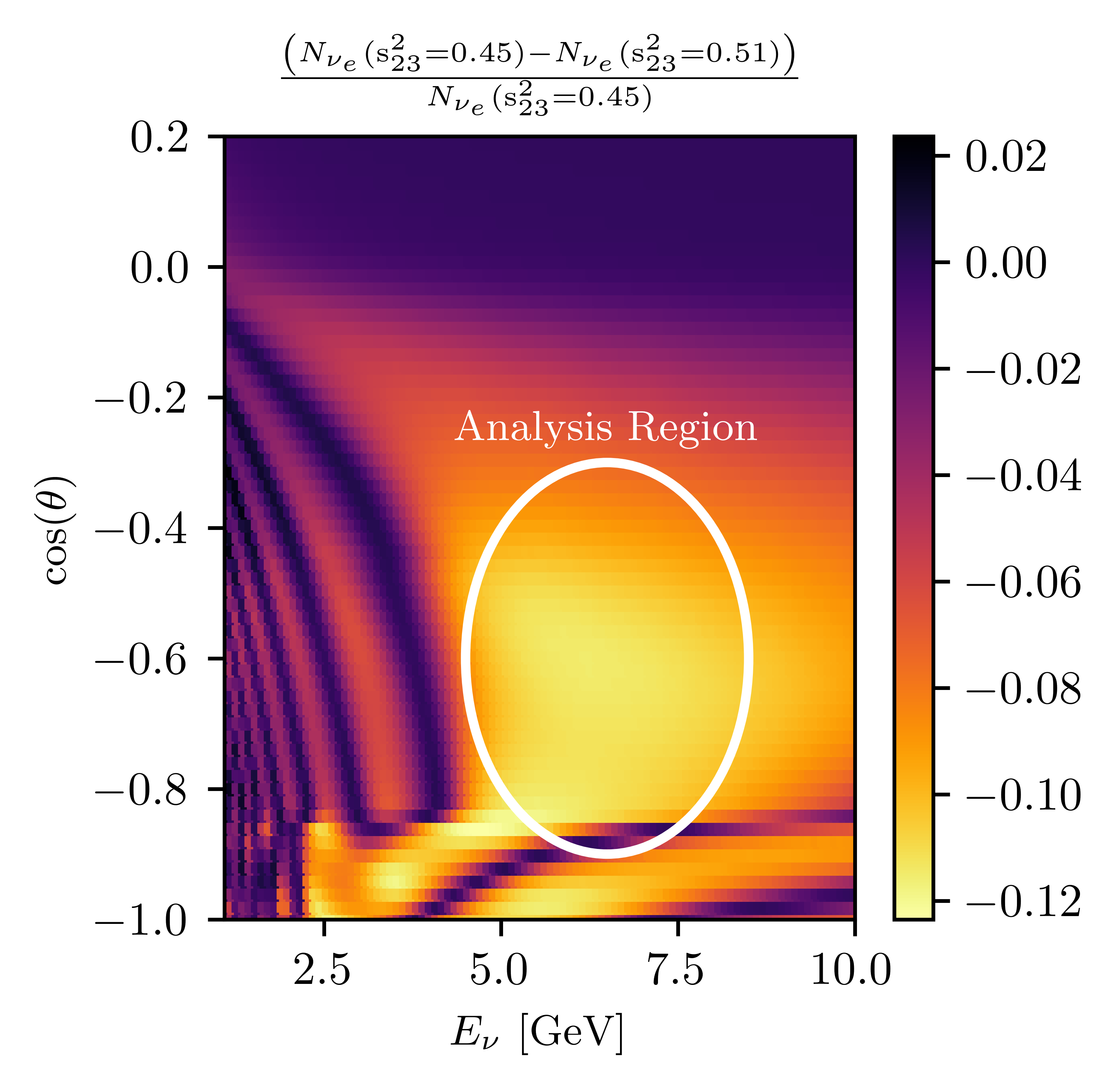}
\vspace{-2\baselineskip}% Reduce image height
\caption{Oscillogram showing the differences between the expected $\nu_e$ counts for $\sin^2\theta_{23} = 0.45$ and $0.51$. The white circle indicates the approximate analysis region used in this study.}
\label{fig:osci}
\end{figure}

For illustrative purposes Fig.~\ref{fig:osci} shows the fractional difference in the expected number of counts in the $\nu_e$ CC channel considering two values of $\sin^2\theta_{23}$. We choose the values $\sin^2\theta_{23} = 0.45$ and $0.51$, which represent the best fit and boundary of the $1\sigma$ experimental uncertainty, respectively. The white circle in Figure~\ref{fig:osci} indicates our approximate analysis region, which focuses on the region most sensitive to changes in $\theta_{23}$. For this example we predict a change of $\sim 12\%$ in total counts in the analysis region. However, distinguishing between these two $\theta_{23}$ values is only possible if the total normalization uncertainty is below 12\%. Assuming a cross section uncertainty of 10\%, we therefore require the neutrino flux uncertainty be below 7\% (root-squared errors) to make such a distinction. As shown in the previous section, Hyper-K will be able to achieve this.

Moving beyond the illustrative example, we now wish to determine Hyper-K's sensitivity to $\theta_{23}$ using the improved primary CR model. To do this we generate $10^6$ sample oscillograms for both the $\nu_e$ and the $\nu_\mu$ CC channels, each including a 10\% cross-section uncertainty. We perform a Poissonian likelihood analysis with an energy-dependent Gaussian nuisance parameter, which captures the cross-section and primary CR uncertainties. Additionally, we maximize the likelihood across energy and angle bins, resulting in the approximate region shown in Figure~\ref{fig:osci}. Finally, $1\sigma$ confidence intervals for $\sin^2\theta_{23}$ are determined keeping all other oscillation parameters constant. 

We run the analysis using three CR uncertainty benchmarks: $15\%$ (present value, as used by Super-K~\cite{Super-Kamiokande:2017yvm}), $7\%$ (achievable value, as demonstrated in the previous section), and $0\%$ (the ideal case). We find that using the present uncertainty value of 15\%, $\sin^2\theta_{23}$ is determined to be $0.45^{+0.61}_{-0.046}$ which does not improve upon the current measurement despite the increased statistics (i.e. it is systematics limited). However, for the 7\% uncertainty case, $0.45^{+0.29}_{-0.022}$ is achievable, 50\%-73\% of the uncertainty in Super-K's $\sin^2\theta_{23}$ measurement~\cite{Super-Kamiokande:2023ahc}. This can be contrasted with the ideal case, which shows that ultimately $\theta_{23}$ can be known to $\leq1\%$ precision if uncertainties in the CR spectrum are negligible. This is a surprising result, given the remaining systematic uncertainties (from the cross section) are 10\%. However, since $\theta_{23}$ causes energy dependent changes in the neutrino spectrum, unlike the systematics, it can be determined to higher precision.

%%%%%%%%%%%%%%%%%%%%%%%%%%%%%%%%%%%%%%%%%%%%%%%%%%%%%%%%%%%%%%%%%%%%%%%%%%%%%%%%%%%%%%%%%%%
%%%%%%%%%%%%%%%%%%%%%%%%%%%%%%%%%%%%%%%%%%%%%%%%%%%%%%%%%%%%%%%%%%%%%%%%%%%%%%%%%%%%%%%%%%%

\section{Conclusion and Outlook}\label{sec:conclusion}

We have shown that next-generation kiloton-scale neutrino detectors, such as Hyper-Kamiokande, can provide a powerful new probe of the primary cosmic ray spectrum. While not designed as cosmic ray experiments, these detectors offer full-sky coverage, long-term stability, and unique sensitivity to low-energy features inaccessible to traditional satellite and balloon missions. 

Hyper-Kamiokande can distinguish between leading cosmic ray models with high significance and reconstruct the primary spectrum to reduce flux uncertainties from $\sim$20\% to $\sim$7\%. This improvement will directly enhance the precision of neutrino oscillation measurements, particularly in parameters such as $\sin^2\theta_{23}$, and sharpen searches for diffuse supernova neutrinos, proton decay, and dark matter.

While the analysis in this paper has focused on the example of Hyper-Kamiokande as a proof of principle, a similar analysis should be possible for JUNO and DUNE. Looking forward with currently-operating experiments, complementary data from detectors like ORCA~\cite{KM3Net:2016zxf} and seasonal muon variations observed by BOREXINO~\cite{Borexino:2018pev}, Super-K~\cite{Super-Kamiokande:2024rwz}, IceCube~\cite{IceCube:2015wro, IceCube:2017qmt, Tilav:2019xmf}, KM3NeT~\cite{KM3NeT:2024buf}, and AUGER~\cite{Gora:2022rkp} offer promising paths to constrain cosmic ray uncertainties further. Our results emphasize that accounting for and reducing cosmic ray systematics will be critical for the next era of precision neutrino physics.

%%%%%%%%%%%%%%%%%%%%%%%%%%%%%%%%%%%%%%%%%%%%%%%%%%%%%%%%%%%%%%%%%%%%%%%%%%%%%%%%%%%%%%%%%%%
%%%%%%%%%%%%%%%%%%%%%%%%%%%%%%%%%%%%%%%%%%%%%%%%%%%%%%%%%%%%%%%%%%%%%%%%%%%%%%%%%%%%%%%%%%%

\section*{Acknowledgements}

We are grateful for helpful discussions with John Beacom and Yi Zhuang. This work was supported by the Australian Research Council through Discovery Project DP220101727 plus the University of Melbourne’s Research Computing Services and the Petascale Campus Initiative. J.L.N was supported by the ARC Centre of Excellence for Dark Matter Particle Physics, CE200100008. L.E.S. acknowledges support from DOE Grant de-sc0010813. This work is based on the ideas and calculations of the authors, plus publicly available information.

%%%%%%%%%%%%%%%%%%%%%%%%%%%%%%%%%%%%%%%%%%%%%%%%%%%%%%%%%%%%%%%%%%%%%%%%%%%%%%%%%%%%%%%%%%%
%%%%%%%%%%%%%%%%%%%%%%%%%%%%%%%%%%%%%%%%%%%%%%%%%%%%%%%%%%%%%%%%%%%%%%%%%%%%%%%%%%%%%%%%%%%
\newpage
\appendix
\counterwithin{figure}{section}
\vspace{0.75cm}
\centerline{\Large {\bf Appendices}}
\vspace{0.75cm}

\section{Interaction models}

In the following we show the uncertainty bands caused by using different hadronic interaction models. To this end we inject either AMS or PAMELA data and generate neutrino spectra using Sybill 2.3~\cite{Riehn:2017mfm, Riehn:2019jet}, EPOS-LHC~\cite{Pierog:2013ria}, QGSJET-II~\cite{Ostapchenko:2010vb}, and DPMJET-III~\cite{Roesler:2000he, Fedynitch:2015kcn} and use those results to rescale the HKKM predictions.

\begin{figure}[t]
\centering
\includegraphics[width=\columnwidth]{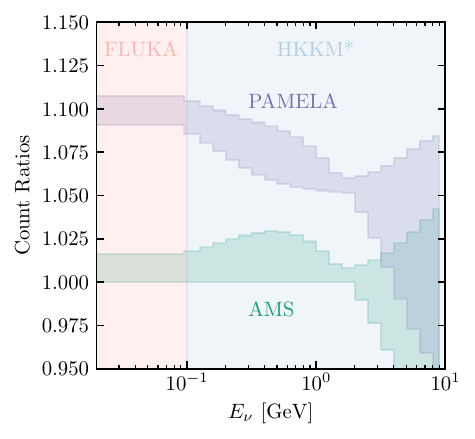}
\vspace{-2\baselineskip}% Reduce image height
\caption{The count ratios including uncertainties from the interaction models. We also denote the relevant energy regions in our simulation. The lack of change at low energies is due to using FLUKA for all simulations.}
\label{fig:model_effects}
\end{figure}

%%%%%%%%%%%%%%%%%%%%%%%%%%%%%%%%%%%%%%%%%%%%%%%%%%%%%%%%%%%%%%%%%%%%%%%%%%%%%%%%%%%%%%%%%%%
%%%%%%%%%%%%%%%%%%%%%%%%%%%%%%%%%%%%%%%%%%%%%%%%%%%%%%%%%%%%%%%%%%%%%%%%%%%%%%%%%%%%%%%%%%%

\clearpage
\bibliography{bibliography}

%%%%%%%%%%%%%%%%%%%%%%%%%%%%%%%%%%%%%%%%%%%%%%%%%%%%%%%%%%%%%%%%%%%%%%%%%%%%%%%%%%%%%%%%%%%
%%%%%%%%%%%%%%%%%%%%%%%%%%%%%%%%%%%%%%%%%%%%%%%%%%%%%%%%%%%%%%%%%%%%%%%%%%%%%%%%%%%%%%%%%%%

\end{document}